\documentclass[a4paper,fleqn]{cas-sc}

\usepackage[authoryear,longnamesfirst]{natbib}

\usepackage{hyperref}

\definecolor{rwth}   {RGB}{  0  84 159}
\definecolor{rwth-75}{RGB}{ 64 127 183}
\definecolor{rwth-50}{RGB}{142 186 229}
\definecolor{rwth-25}{RGB}{199 221 242}
\definecolor{rwth-10}{RGB}{232 241 250}

\definecolor{black}   {RGB}{  0   0   0}
\definecolor{black-75}{RGB}{100 101 103}
\definecolor{black-50}{RGB}{156 158 159}
\definecolor{black-25}{RGB}{207 209 210}
\definecolor{black-10}{RGB}{236 237 237}

\definecolor{magenta}   {RGB}{227   0 102}
\definecolor{magenta-75}{RGB}{233  96 136}
\definecolor{magenta-50}{RGB}{241 158 177}
\definecolor{magenta-25}{RGB}{249 210 218}
\definecolor{magenta-10}{RGB}{253 238 240}

\definecolor{yellow}   {RGB}{255 237   0}
\definecolor{yellow-75}{RGB}{255 240  85}
\definecolor{yellow-50}{RGB}{255 245 155}
\definecolor{yellow-25}{RGB}{255 250 209}
\definecolor{yellow-10}{RGB}{255 253 238}

\definecolor{petrol}   {RGB}{  0  97 101}
\definecolor{petrol-75}{RGB}{ 45 127 131}
\definecolor{petrol-50}{RGB}{125 164 167}
\definecolor{petrol-25}{RGB}{191 208 209}
\definecolor{petrol-10}{RGB}{230 236 236}

\definecolor{turkis}   {RGB}{  0 152 161}
\definecolor{turkis-75}{RGB}{  0 177 183}
\definecolor{turkis-50}{RGB}{137 204 207}
\definecolor{turkis-25}{RGB}{202 231 231}
\definecolor{turkis-10}{RGB}{235 246 246}

\definecolor{grun}   {RGB}{ 87 171  39}
\definecolor{grun-75}{RGB}{141 192  96}
\definecolor{grun-50}{RGB}{184 214 152}
\definecolor{grun-25}{RGB}{221 235 206}
\definecolor{grun-10}{RGB}{242 247 236}

\definecolor{maigrun}   {RGB}{189 205   0}
\definecolor{maigrun-75}{RGB}{208 217  92}
\definecolor{maigrun-50}{RGB}{224 230 154}
\definecolor{maigrun-25}{RGB}{240 243 208}
\definecolor{maigrun-10}{RGB}{249 250 237}

\definecolor{orange}   {RGB}{246 168   0}
\definecolor{orange-75}{RGB}{250 190  80}
\definecolor{orange-50}{RGB}{253 212 143}
\definecolor{orange-25}{RGB}{254 234 201}
\definecolor{orange-10}{RGB}{255 247 234}

\definecolor{rot}   {RGB}{204   7  30}
\definecolor{rot-75}{RGB}{216  92  65}
\definecolor{rot-50}{RGB}{230 150 121}
\definecolor{rot-25}{RGB}{243 205 187}
\definecolor{rot-10}{RGB}{250 235 227}

\definecolor{bordeaux}   {RGB}{161  16  53}
\definecolor{bordeaux-75}{RGB}{182  82  86}
\definecolor{bordeaux-50}{RGB}{205 139 135}
\definecolor{bordeaux-25}{RGB}{229 197 192}
\definecolor{bordeaux-10}{RGB}{245 232 229}

\definecolor{violett}   {RGB}{ 97  33  88}
\definecolor{violett-75}{RGB}{131  78 117}
\definecolor{violett-50}{RGB}{168 133 158}
\definecolor{violett-25}{RGB}{210 192 205}
\definecolor{violett-10}{RGB}{237 229 234}

\definecolor{lila}   {RGB}{122 111 172}
\definecolor{lila-75}{RGB}{155 145 193}
\definecolor{lila-50}{RGB}{188 181 215}
\definecolor{lila-25}{RGB}{222 218 235}
\definecolor{lila-10}{RGB}{242 240 247}

\hypersetup{
	colorlinks=true,
	linkcolor=rwth,
	anchorcolor=black,
	citecolor=rwth,
	filecolor=rwth,
	menucolor=rwth,
	urlcolor=rwth
}

\usepackage{xcolor}
\usepackage{siunitx}
\usepackage{subcaption}
\usepackage{nicefrac}
\usepackage{todonotes}
\graphicspath{{./figs/}}

\newcommand{\vect}[1]{\boldsymbol{\mathit{#1}}}
\newcommand{\tens}[1]{\boldsymbol{\mathrm{#1}}}

\newcommand{\npfrac}[2]{\nicefrac{\partial #1}{\partial #2}}

\newcommand{\ICdev}{\mathrm{I}_{\tens{\bar C}}}
\newcommand{\IICdev}{\mathrm{II}_{\tens{\bar C}}}
\newcommand{\IIICdev}{\mathrm{III}_{\tens{\bar C}}}
\newcommand{\pfrac}[2]{\frac{\partial #1}{\partial #2}}

\newcommand{\IC}{\mathrm{I}_{\tens{C}}}
\newcommand{\IIC}{\mathrm{II}_{\tens{C}}}
\newcommand{\IIIC}{\mathrm{III}_{\tens{C}}}

\DeclareMathOperator{\tr}{tr}

\begin{document}
\let\WriteBookmarks\relax
\def\floatpagepagefraction{1}
\def\textpagefraction{.001}
\shorttitle{Rediscovering Hyperelasticity by Deep Symbolic Regression}
\shortauthors{R. Abdusalamov and M. Itskov}

\title [mode = title]{Rediscovering Hyperelasticity by Deep Symbolic Regression}

\author[1]{Rasul Abdusalamov}[type=editor,
auid=000,bioid=1,
orcid=0000-0003-4988-4794]
\cormark[1]
\fnmark[1]
\ead{abdusalamov@km.rwth-aachen.de}
\ead[url]{www.km.rwth-aachen.de}

\affiliation[1]{organization={Department of Continuum Mechanics, RWTH Aachen University},
	addressline={Eilfschornsteinstr. 18}, 
	city={Aachen},
	postcode={52062}, 
	state={NRW},
	country={Germany}}

\author[1,2]{Mikhail Itskov}[]

\fnmark[2]
\ead{itskov@km.rwth-aachen.de}
\ead[URL]{www.km.rwth-aachen.de}

\affiliation[2]{organization={School of Computer Science \& Mathematics, Keele University},
	postcode={ST5 5BG}, 
	postcodesep={,}, 
	city={Staffordshire},
	country={United Kingdom}}

\cortext[cor1]{Corresponding author}

\begin{abstract}
	Accurate hyperelastic material modeling of elastomers under multiaxial loading still remain a research challenge. This work employs deep symbolic regression as an interpretable machine learning approach to discover novel strain energy functions directly from experimental results, with a specific focus on the classical Treloar and Kawabata data sets for vulcanized rubber. The proposed approach circumvents traditional human model selection biases by exploring possible functional forms of strain energy functions expressed in terms of both the first and second principal invariants of the right Cauchy-Green tensor. The resulting models exhibit high predictive accuracy for various deformation modes, including uniaxial and equibiaxial tension, pure shear, as well as a general biaxial loading. This  underscores the potential of deep symbolic regression in advancing hyperelastic material modeling and highlights the importance of both invariants in capturing the complex behaviors of rubber-like materials.
\end{abstract}

\begin{highlights}
	\item Proposed novel hyperelastic material models discovered through deep symbolic regression from experimental data eliminating human bias in model selection
	\item Accurately described classical Treloar and Kawabata data sets with interpretable strain energy functions requiring few material parameters 
	\item Confirmed that utilizing both the first and second invariants are crucial for accurately capturing complex behavior of rubber-like materials under diverse loading conditions
\end{highlights}

\maketitle

\section{Introduction}
To predict the mechanical behavior of materials under various loading conditions especially by finite elements (FE), it is essential to employ constitutive models of the highest possible accuracy. This is particularly true in the case of elastomeric materials subjected to large deformations. In literature, numerous hyperelastic models have been proposed for elastomers. However, not all of them are capable of reproducing the complete mechanical behavior across different loading types and for a wide range of different rubber-like materials. This situation often poses a significant challenge to engineers in selecting the most appropriate model with minimal number of material parameters. A comprehensive review of the literature pertaining to hyperelastic materials models and their functionality is provided in a series of papers, including those by \cite{Marckmann2006, Xiang2020, Melly2021, Dal2021, he2022comparative, ricker2023systematic} and numerous references therein. 
In recent years, a considerable number of additional material models have been reported, with a notable emphasis on highly specialized materials and particular effects. For example, on micro-mechanically based constitutive models \citep{itskov2016rubber,Khiem2017, MIRZAPOUR2023112299}, interpolation-based approaches for phenomenological constitutive models \citep{MENG2021101485} or even strain-mode-dependent concepts \citep{mahnken2022strain} have been proposed. Nevertheless, a critical evaluation of the existing modeling approaches reveals several drawbacks. First, the calibration of material models is often challenging due to dependence on multiple material parameters. Second, many models exhibit constrained predictive accuracy in specific loading scenarios, attributable to the nature of input data and assumptions made. Third, implementing each model as a user material model in commercial FE solvers necessitates a substantial investment of time. Finally, all of the aforementioned models have been developed by humans, which introduces a potential bias and consequently results in a limitation of predictive capabilities. For these reasons, a more sophisticated approach is necessary, capable of generating specific material models while maintaining reasonable computation time without the need for extensive expert knowledge. Furthermore, the approach must be implementable for practical use in industrial applications. \\
Some of these limitations can be circumvented by employing data-driven techniques. With the advancement of machine learning, numerous data-driven modeling approaches have recently been developed for constitutive modeling of materials. Detailed reviews of such approaches can be found e.g. in \cite{herrmann2024deep}  and \cite{fuhg2024review}. They include physics-augmented neural networks \citep{kalina2023fe}, polyconvex anisotropic hyperelasticity with neural networks \citep{klein2022polyconvex}, discovery of material models using sparse regression \citep{flaschel2023automated}, model-free approaches \citep{kirchdoerfer2016data}, constitutive artificial neural networks \citep{Linka2020}, and constitutive Kolmogorov–Arnold networks \citep{abdolazizi2025constitutive}. Many of these models introduce theoretical concepts of materials theory or thermodynamics into the computational framework and focus on invariant or strain-based approaches. Despite their advantages, these methods are not without serious limitations. While artificial neural networks offer significant benefits in terms of model discovery, they are often regarded as "black boxes", which poses a substantial challenge to interpretation and usage in industrial applications. Furthermore, the training of such models necessitates substantial computational effort, and the high complexity of resulting models can hinder further applications, such as FE simulations. In addressing these limitations, efforts have been made to enhance interpretability and to significantly reduce the complexity of the neural networks (see, e.g. \citep{linka2023new}). An alternative approach could be to employ a conventional nonlinear optimization scheme that would not necessitate a complex neural network architecture. Furthermore, many of the presented methods do not lead to the discovery of novel material models. The identification of a most appropriate combination of already known terms from a given set of functions closely aligned with the underlying data remains the primary objective. A further drawback is that many approaches combine redundant inputs such as invariants and principal stretches (see, e.g. \cite{abdolazizi2025constitutive}). \\
An alternative method that has been demonstrated to overcome many of the disadvantages of neural networks is symbolic regression (SR). Symbolic regression is a relatively novel regression method that belongs to the class of interpretable machine learning algorithms. It determines a mathematical expression by searching a solution space where the best-fitting expression structure is identified for a given data set \citep{Augusto2000}. Accordingly, an expression that is optimal in terms of simplicity and accuracy with respect to the data set is formulated. The principal benefit of this approach is that it identifies an analytical model while reducing the effect of human bias. Recent applications include the development of interpretable hyperelastic material models \citep{Abdusalamov2023}, plasticity models \citep{bomarito2021development}, modeling the Mullins effect \citep{ABDUSALAMOV2024}, and learning  implicit yield surface models using uncertainty quantification \citep{birky2025learning}. Furthermore, a novel method for constitutive law discovery that relies on formal grammars and shares notable similarities with a symbolic regression approach has recently been proposed \citep{kissas2024language}. \\
Extending the work presented in \cite{ABDUSALAMOV2024} we employ here deep symbolic regression as an interpretable machine learning approach to discover novel strain energy functions directly from experimental data. In particular, we focus on the classical Treloar and Kawabata data sets for vulcanized rubber. The proposed approach circumvents traditional human model selection biases. The resulting models demonstrate high levels of predictive accuracy across various deformation modes, including uniaxial tension, pure shear and biaxial tension. \\ 
The structure of the paper is as follows: \autoref{sec:Methodlogy} discusses the proposed methodology, including an overview of deep symbolic regression embedded into a continuum mechanical framework. \autoref{sec:ResultsAndDiscussion} presents the results for the newly discovered strain energy functions for the Treloar and Kawabata data sets. Additionally, the robustness of the presented approach is evaluated with respect to noise. Furthermore, a stretch-based approach is discussed. Finally, a brief conclusion highlights the main aspects of this work in \autoref{sec:Conclusion}.

\section{Methodology}
\label{sec:Methodlogy}
\subsection{Continuum Mechanical Framework}
A strain energy function $\Psi(\tens C)$ of an isotropic hyperelastic material can be expressed in terms of the principal invariants $\IC, \IIC$ and $\IIIC$ of the right Cauchy-Green tensor $\tens{C}=\tens{F}^\text{T}\tens{F}$ , where $\tens{F}$ denotes the deformation gradient. Consequently, the first Piola-Kirchhoff stress tensor $\tens{P}$ can be expressed as follows:
\begin{align}
	\label{equ:S}
	\tens{P}
	= 2 \tens{F}\frac{\partial \Psi(\tens{C})}{\partial \tens{C}} = 2\left[\left({\color{rwth} \pfrac{\Psi}{\IC}} + \IC {\color{rwth} \pfrac{\Psi}{\IIC}} \right) \mathbf{F} - {\color{rwth} \pfrac{\Psi}{\IIC}} \tens{FC} + \IIIC {\color{rwth} \pfrac{\Psi}{\IIIC}} \tens{F}^{-\text{T}}\right].
\end{align}
It is important to highlight that the impact of the material is solely determined by the terms marked {\color{rwth} blue}. The invariants $\IC$, $\IIC$ and $\IIIC$ of $\tens C $ are given by
\begin{align}
	\IC=\tr\tens{C} \,, \quad  \IIC=\frac{1}{2}\left[\left(\tr\tens{C}\right)^2-\tr\left(\tens{C}^2\right)\right] \,,  \quad \text{and} \quad   \IIIC=\det\tens{C} \, .
\end{align}
$\Psi(\tens{C})$ should satisfy the conditions of the energy and stress-free natural state at $\mathbf{F}=\mathbf{I}$ given by
\begin{align}\label{stress-free}
	\Psi(\mathbf{I})=0 \quad \text{and} \quad   \frac{\partial \Psi(\tens{C})}{\partial \tens{C}} \Bigg|_{\tens{C}=\tens{I}} =\tens{0} \, .
\end{align}
In the case of nearly incompressible behavior, it is common to multiplicatively decompose the deformation gradient into a volumetric $\tens{\hat F} = J\tens{I}$ and an isochoric part $\tens{\bar F}=J^{-1/3}\tens{F}$, where $J= \det\tens{F} = \sqrt{\IIIC}$ (see \citep{richter1948isotrope} for further details). Consequently, the principal invariants of the isochoric right Cauchy-Green tensor $\tens{\bar C}=\tens{\bar F}^{\text{T}}\tens{\bar F}$ take the form
\begin{align}
	\ICdev&= J^{-2/3}\mathrm{I}_{\tens{ C}} \, , \quad \IICdev = J^{-4/3}\mathrm{II}_{\tens{ C}} \quad \text{and} \quad \IIICdev = 1 \, .
\end{align}
Accordingly, the first Piola-Kirchhoff stress tensor can be expressed as
\begin{align} \label{P-v-i-split}
	\tens{P} = 2 \left( {\color{rwth} \pfrac{\Psi}{\ICdev}}  + \ICdev {\color{rwth} \pfrac{\Psi}{\IICdev}}  \right) J^{-2/3} \mathbf{F} - 2 {\color{rwth} \pfrac{ \Psi}{\IICdev} } J^{-4/3}  \tens{FC} + J \left({\color{rwth} \pfrac{ \Psi}{J}} - \frac{2}{3J} {\color{rwth} \pfrac{ \Psi}{\ICdev}}  \ICdev   - \frac{4}{3J} {\color{rwth} \pfrac{ \Psi}{ \IICdev} } \IICdev\right) \tens{F}^{-\text{T}}.
\end{align}
In the case of ideally incompressible materials, characterized by the constraint $J=1$, the constitutive equation will take the following form:
\begin{align}
	\mathbf{P} =2 \tens{F}\frac{\partial \Psi(\tens{C})}{\partial \tens{C}} -p \tens{F}^{-\text{T}} =2\left[\left( {\color{rwth} \pfrac{ \Psi}{ \IC}} + \IC {\color{rwth} \frac{ \Psi}{\IIC}} \right) \mathbf{F}- {\color{rwth} \pfrac{ \Psi}{ \IIC} } \tens{FC} \right] \label{P-incompr}
	-p \tens{F}^{-\text{T}} \, ,
\end{align}
where $p$ denotes an arbitrary parameter related to the hydrostatic pressure. While the proposed formulations use strain invariants, it is also feasible to express $\Psi$ in terms of the principal stretches $\lambda_i (i=1,2,3)$. Then, in the case of distinct $\lambda_i$ we can write 
\begin{align}
\tens{P}=\sum_{i=1}^3 \frac{\partial \Psi\left(\lambda_1, \lambda_2, \lambda_3\right)}{\partial \lambda_i} \vect{n}_i \otimes \vect{N}_i ,
\end{align}
where $\vect{n}_i$ and $N_i$ are unit eigenvectors of $\tens{b}= \tens{ F}\tens{ F}^{\text{T}}$ and $\tens{ C}$, respectively.\\
According to the Valanis-Landel concept, the strain-energy density function $\Psi$ of a hyperelastic material can be expressed in terms of a continuously differentiable function $\omega$ as follows \citep{valanis1967strain}
\begin{align}
	\Psi(\lambda_1, \lambda_2, \lambda_3) = \omega(\lambda_{1}) + \omega(\lambda_{2}) + \omega(\lambda_{3}) \, .\label{eq:ValanisLandelAssumption}
\end{align}
The above mentioned conditions of the zero energy and stress free reference state require that
\begin{align}
	\omega(1) = 0 \quad \text{and} \quad \omega'(1) = 0 \, .
\end{align}
Within the proposed framework of deep symbolic regression, both the invariant and stretch based formulation can be used to identify strain energy functions describing a specified data set. The well-known Ogden model represents a special case of \eqref{eq:ValanisLandelAssumption} with  
\begin{align}
\omega(\lambda) = \sum_{i=1}^{N} \frac{\mu_{i}}{\alpha_{i}} \left(\lambda^{\alpha_{i}} - 1\right) \, .
\end{align}
It gained wide acceptance in industrial FE software and has demonstrated excellent performance \citep{ogden1972large}.

\subsection{Deep Symbolic Regression}
\begin{figure}[h!]
	\centering
	\includegraphics[width=\textwidth]{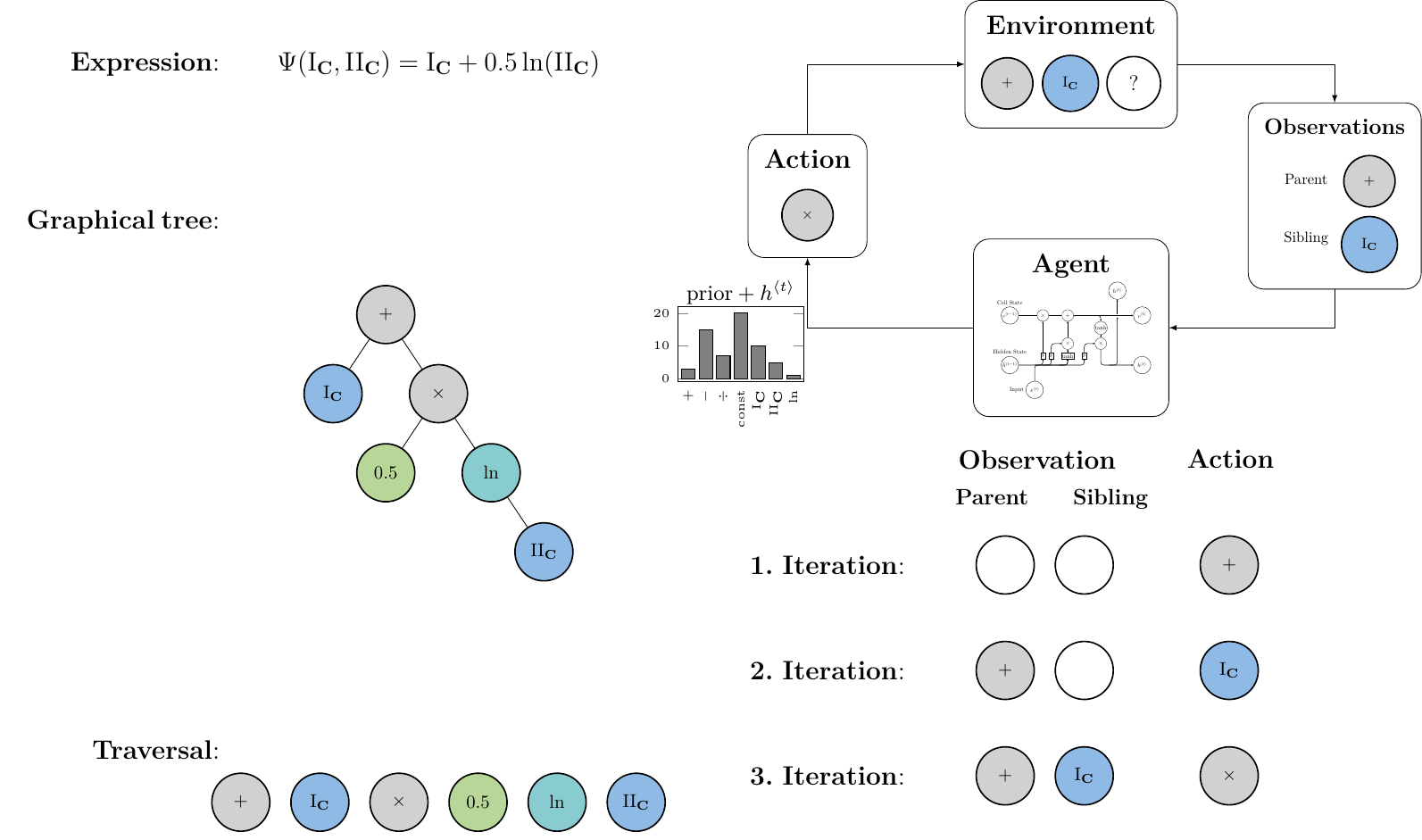}
	\caption{Expression sampling process of the deep symbolic regression framework applied to derive the strain energy function $\Psi(\IC, \IIC) = \IC \,+\, 0.5\ln(\IIC)$. The RNN samples tokens given by operations, functions, constants  inputs to build an expression tree starting from a root token. The rewards are computed using the NRMSE to train the RNN using a risk-seeking policy gradient.}
	\label{fig:DSOProcess}
\end{figure}

The deep symbolic regression package by \cite{Petersen2021} utilizes a recurrent neural network (RNN) to predict a mathematical expression based on a sampled distribution through a risk-seeking policy gradient. The framework is principally based on a reinforcement learning (RL) approach. Each expression tree is transformed into a sequence, referred to as a traversal, which corresponds to the environment of the reinforcement learning task (see \autoref{fig:DSOProcess}). The node values of the traversal referred to as tokens, represent either operations, functions, constants, or arguments. The recurrent neural network, serving as the agent is trained iteratively on a hierarchical input containing information about the entire expression tree. The traversal is decomposed into observations about siblings and parents fed directly into the RNN. The next element of the traversal, corresponding to the action, is sampled based on a probability distribution function. A reward function is formulated based on the performance of the sampled expression on the given data set.
The detailed process of the RL algorithm is illustrated for example by the strain energy function $\Psi(\IC, \IIC) = \IC + 0.5\ln(\IIC)$ (see \autoref{fig:DSOProcess}) and proceeds as follows:
\begin{enumerate}
	\item In each epoch, a batch of expressions is sampled according to the following steps:
	\begin{enumerate}[i.]
			\item In the initialization phase the sampling of an initial token (the root of the expression tree) from a library containing all the necessary operations, functions, constants, and arguments takes place. The sampling of the token is derived from a predefined probability distribution. This initial step does not specify any information about parent or sibling relationships. In the illustrated example, the first sampled token in the first iteration represents the addition operator ${\color{bordeaux}+}$.
			\item Sampling of subsequent tokens requires update of observations based on the previous token, the updating of the weights of the RNN, and the sampling of the next token. An advantage of this approach is that search space constraints can be incorporated directly into the sampling process, which can be achieved by introducing a prior into the probability distribution function. One such prior constraint on the search space is that all children of an operator cannot be constants, given that they would otherwise be reduced to another constant. For instance, in the second iteration of the subsequent sampling step, the addition operator is designated as the parent with an arity of two, since there are no siblings and no additional information is specified. Thus, the first input ${\color{rwth} \IC}$ is the next sampled token. 
			\item This iterative process continues as long as to all nodes in the tree a terminal status has been assigned, which is either a constant or an input variable. In this manner, each token within the expression for $\Psi(\IC, \IIC)$ is sampled stepwise until an expression is determined.
		\end{enumerate}
	\item Based on the generation of expressions, the reward is calculated using the normalized root mean square error (NRMSE). A risk-seeking policy gradient is implemented with the objective of maximizing the performance of a specified fraction of the best samples. Consequently, the best case performance is prioritized, albeit at the potential expense of lower worst case and average performance.
\end{enumerate}
Furthermore, the \texttt{DSO} package offers a constant optimization option. Despite the increased time requirement and a certain risk of overfitting, this approach allows for a significantly higher rate of the expression recovery. Once the prior has been sampled, the corresponding symbolic expression is instantiated and evaluated. In this work, the following basis functions for the strain energy are applied: ["\textsf{add}", "\textsf{sub}",  "\textsf{n2}", "\textsf{mul}", "\textsf{div}", "\textsf{sqrt}", "\textsf{exp}", "\textsf{log}"]. The normalization for the strain energy, as given by condition \eqref{stress-free}$_{1}$ can easily be satisfied by correction of the resulting expression by a constant, which has been omitted in the following. Condition \eqref{stress-free}$_{2}$ is fulfilled automatically by including the point $\tens{P} = \tens{0}$ at $\tens{F} = \tens{I}$ into the set of the data used for the search of the mathematical expression of the strain energy function.
\section{Results and Discussion}
\label{sec:ResultsAndDiscussion}
In the following we apply deep symbolic regression to experimental data by \cite{treloar1944stress} and \cite{kawabata1981experimental} from uniaxial and biaxial tests of a vulcanized rubber. Fitting these classical data sets remains a significant challenge in hyperelastic material modeling. It is also a critical step towards a deeper understanding of the complex behavior of rubber-like materials under diverse deformation modes.
\subsection{Multi-Axial Loading of Vulcanized Rubber}
\label{sec:MutliAxialVulRub}
\begin{figure}[h!]
	\centering
	\includegraphics[width=0.6\textwidth]{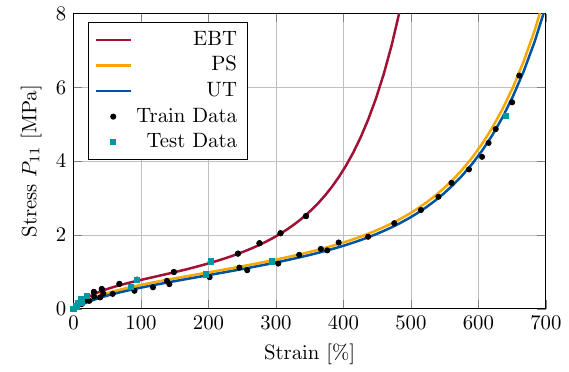} 
	\caption[Best Fit for Treloar Data of Novel Model Discovered Using \texttt{DSO}]{Best fit for the Treloar data set of novel model discovered through \texttt{DSO}. The stress-strain responses for UT, PS and EBT are generated from the strain energy function given in \autoref{equ:TreloarBestFitDSO}.}
	\label{fig:TreloarFit_no_noise}
\end{figure}

\begin{figure}[p]
	\centering
		\begin{subfigure}{0.5\textwidth}
			{\includegraphics[width=1.2\textwidth]{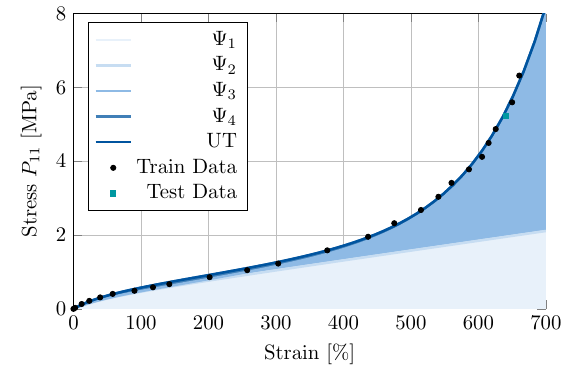}}
			\caption{}
			\label{fig:Treloar_Terms_UT}
			\end{subfigure}\\
		\begin{subfigure}{0.5\textwidth}
			{\includegraphics[width=1.2\textwidth]{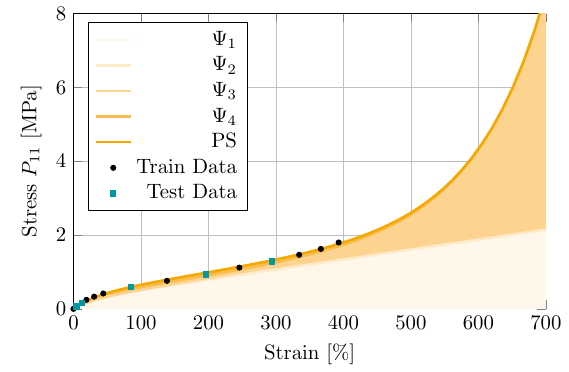}}
			\caption{}
			\label{fig:Treloar_Terms_PS}
			\end{subfigure}\\
		\begin{subfigure}{0.5\textwidth}
			{\includegraphics[width=1.2\textwidth]{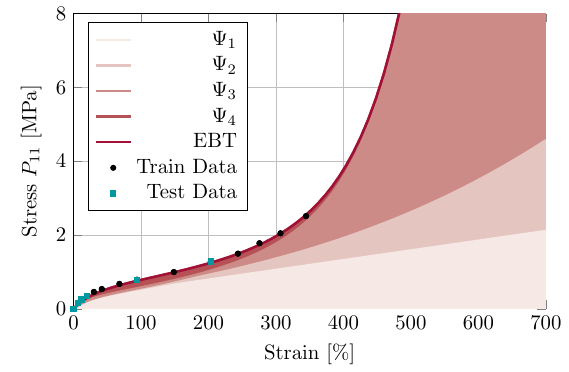}}
			\caption{}
			\label{fig:Treloar_Terms_EBT}
			\end{subfigure}	
	\caption{Visualization of the contributions of each term $\Psi_{i}$ for $i=1,\dots,4$ in the strain energy \eqref{equ:TreloarBestFitDSO}. The responses are shown for UT, PS and EBT for the strain range from $0\SI{}{\percent}$ to $700\SI{}{\percent}$.}
	\label{fig:TreloarContributions}
\end{figure}

\begin{figure}[p]
	\centering
	\begin{subfigure}{0.5\textwidth}
			\includegraphics[width=1.2\textwidth]{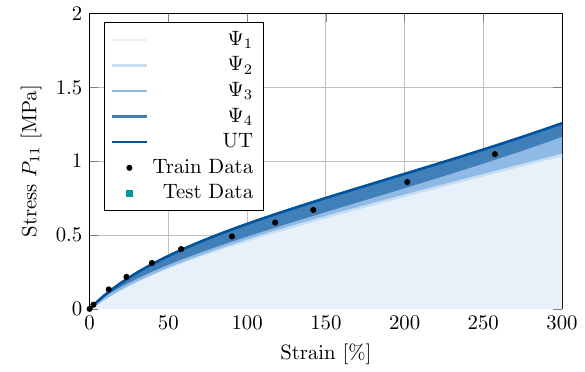} 
			\caption{}
			\label{fig:Treloar_Terms_UT_zoom}
		\end{subfigure} \\
	\begin{subfigure}{0.5\textwidth}
			\includegraphics[width=1.2\textwidth]{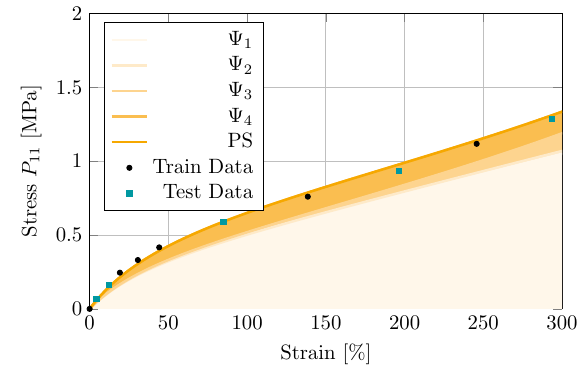}
			\caption{}
			\label{fig:Treloar_Terms_PS_zoom}
		\end{subfigure} \\
	\begin{subfigure}{0.5\textwidth}
			\includegraphics[width=1.2\textwidth]{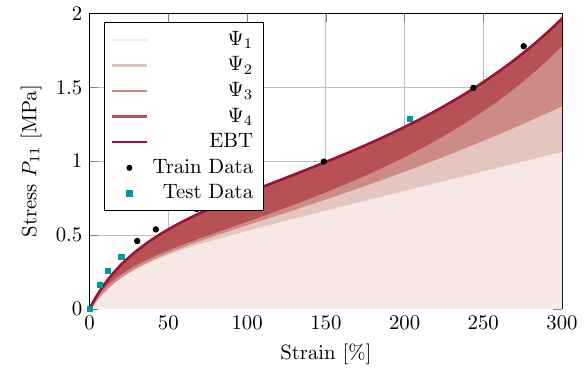}
			\caption{}
			\label{fig:Treloar_Terms_EBT_zoom}
		\end{subfigure}
	\\
	\caption[Contributions of Terms in Strain Energy for Strains up to $\SI{300}{\percent}$]{Visualization of the contributions of each term $\Psi_{i}$ for $i=1,\dots,4$ in the strain energy \eqref{equ:TreloarBestFitDSO}. The responses are shown for UT, PS and EBT for the strain range from $0\SI{}{\percent}$ to $300\SI{}{\percent}$.}
	\label{fig:TreloarContributionsZoom}
\end{figure}

First the \texttt{DSO} package is employed to identify an optimal strain energy function for describing the Treloar data set from pure shear (PS), uniaxial (UT) and equibiaxial tension (EBT) tests. This data set has become one of the most well-known and frequently utilized benchmark tests for hyperelastic models, see e.g. \cite{ricker2023systematic, he2022comparative, Marckmann2006} as well as references therein. All the models proposed in literature have been motivated by human bias to describe the underlying data. The question remains as to whether alternative unbiased model structures exist that could describe the data set more effectively with fewer material parameters.\\
In the following, the data is divided into a training ($80\SI{}{\percent}$) and a test set ($20\SI{}{\percent}$). Subsequently, the efficacy of the identified strain energy functions is evaluated in comparison to these best established models, as referenced in \citep{Marckmann2006, ricker2023systematic, he2022comparative}. To identify an appropriate material model, it is necessary to fit all three curves by \cite{treloar1944stress} simultaneously. The data set contains 14 data points for the EBT and PS, respectively, and 25 data points for the UT response. Given the significant difference in the number of data points for EBT and UT, a straightforward approach is to duplicate the data set and assign a greater weight to the EBT response. This procedure is applied to all fits for all models compared. The best performing strain energy $\Psi_{\text{T}}$ discovered is given by
\begin{align}
	\Psi_{\text{T}} =  & \underbrace{0.13\IC}_{\Psi_{1}} + \underbrace{\SI{2.4011865971618606e-3}{} \IIC}_{\Psi_{2}}  + \underbrace{\SI{001.9961020686934149e-3 }{}\exp{\sqrt{\IC}}}_{\Psi_{3}} + \underbrace{\SI{2.7563861123868053e-2}{} \left(\left(\ln{\IC}\right)^{2} + \left(\ln{\IIC}\right)^{2}\right)  }_{\Psi_{4}} \, . \label{equ:TreloarBestFitDSO}
\end{align}
It demonstrates a high degree of predictive power, as evidenced by an $R^{2}$ score of $\SI{97.31578576913696}{\percent}$. \autoref{fig:TreloarFit_no_noise} illustrates the UT, PS, and EBT responses of this hyperelastic model.
Interestingly, strain energy function \eqref{equ:TreloarBestFitDSO} can be split up into four additive terms. Separate contributions of these terms under UT, PS and EBT are  illustrated in \autoref{fig:TreloarContributions} and \autoref{fig:TreloarContributionsZoom} for the whole and moderate range of deformations, respectively. The first basic term $\Psi_{1}$ represents the neo-Hookean model resulting from Gaussian chain statistics of polymer chains and is thus physically motivated. This term plays a pivotal role in the UT and PS responses, particularly within the range of small strains up to $300\SI{}{\percent}$. $\Psi_{2}$ belongs to the well-known Mooney-Rivlin model and appears to be particularly significant for strains exceeding $300\SI{}{\percent}$, resulting in a markedly more rigid response under EBT. The contribution of $\Psi_{3}$ is of particular significance for all loading cases for the strain range over $300\SI{}{\percent}$. The last term $\Psi_4$ becomes especially important under EBT for the strain range between $0\SI{}{\percent}$ and $300\SI{}{\percent}$. \\
\cite{Rivlin1951} observed in their experiments with a comparable rubber that $\npfrac{\Psi}{\IC}$ is independent of both $\IC$ and $\IIC$, while $\npfrac{\Psi}{\IIC}$ is independent of $\IC$  and decreases with increasing $\IIC$. In light of these observations, they proposed a strain energy function of the form:
\begin{align}
		\Psi_{\text{T}} =c_{10}\left(\IC -3 \right) + \Phi\left(\IIC - 3\right) \, ,
\end{align}
where $c_{10}$ represents a constant, while $\Phi$ denotes a differentiable concave function. Based on this work, \cite{gent1958} proposed $\Phi$ in the form
\begin{align}
	\Phi = c_{01} \ln \left(\frac{\IIC}{3}\right) \, ,
\end{align}
where $c_{01}$ is a constant. The strain energy function \eqref{equ:TreloarBestFitDSO} appears to reflect this underlying logic, as evidenced by the term $\Psi_{4}$. However, $\Psi_4$ is markedly more intricate and nonlinear in character than the classical models proposed for $\Phi$. It is a considerable challenge for human intuition to successfully identify such a contribution. This term has a significant impact on the EBT response, while its influence under UT and PS is negligible. \\
The found material model \eqref{equ:TreloarBestFitDSO} is a more accurate representation of the data set than many traditional hyperelastic models and requires only four material constants. \cite{Marckmann2006} provided a summary and comparative analysis of twenty different models with respect to their ability to fit the experimental data by \cite{treloar1944stress}. It was observed that the extended-tube model with only four material parameters exhibits the best performance characteristics among all the models under consideration \citep{kaliske1999extended}. Furthermore, the non-hyperelastic Shariff \citep{Shariff2000} and the unit sphere models \citep{MIEHE20042617} demonstrated a high level of accuracy.
The three-terms Ogden model is extensively utilized in FE simulations and accurately describes the underlying experimental data. However, determination of its six material parameters necessitates a substantial experimental data set for precise fitting. 
\begin{figure}[p!]
	\centering
	\begin{subfigure}{0.5\textwidth}
			{\includegraphics[width=1.2\textwidth]{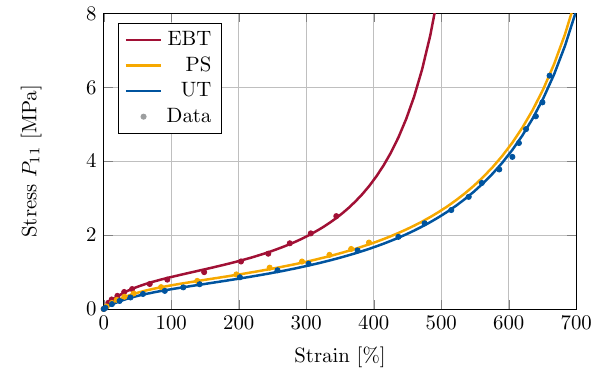}}
			\caption{}
			\label{fig:Treloar_Comparison_Model_1}
		\end{subfigure}\\
	\begin{subfigure}{0.5\textwidth}
			{\includegraphics[width=1.2\textwidth]{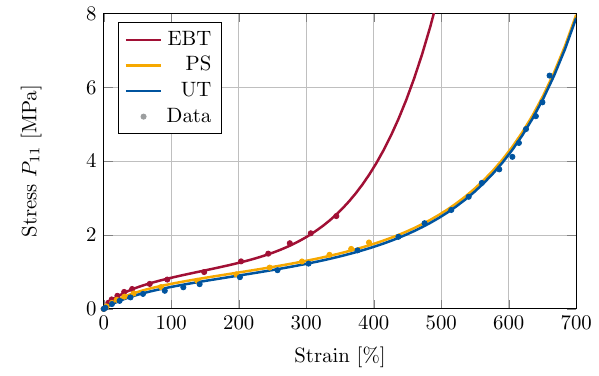}}
			\caption{}
			\label{fig:Treloar_Comparison_Model_2}
		\end{subfigure}\\
	\begin{subfigure}{0.5\textwidth}
			{\includegraphics[width=1.2\textwidth]{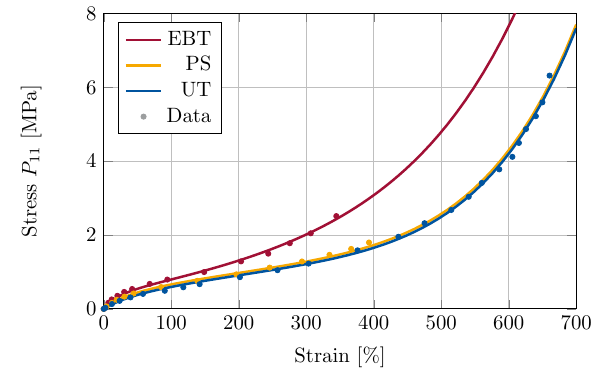}}
			\caption{}
			\label{fig:Treloar_Comparison_Model_3}
		\end{subfigure}	
	\caption[]{Best fit for the Treloar data set using (\subref{fig:Treloar_Comparison_Model_1}) the extended tube model ($R^2 = \SI{96.56108483739724}{\percent}$), (\subref{fig:Treloar_Comparison_Model_2}) the non-hyperelastic Shariff model ($R^2 = \SI{96.37595636432934}{\percent}$) and (\subref{fig:Treloar_Comparison_Model_3}) the stretch-based Ogden model ($R^2 = \SI{95.58273523135516}{\percent}$) for UT, PS and EBT. The used material parameters are listed in \autoref{tab:ExtendedTubeTreloarMaterialParam}, \autoref{tab:ShariffTreloarMaterialParam} and \autoref{tab:OgdenTreloarMaterialParam}, respectively.}
	\label{fig:Treloar_Comparison}
\end{figure}
\begin{table}[h]
	\centering
	\caption{Comparison of the proposed model with existing approaches with respect to the number of material constants and the $R^{2}$ score for the Treloar data for each model.}
	\label{tab:ModelComparison}
	\renewcommand{\arraystretch}{1.5}
	\begin{tabular}{ccc}
			\toprule
			Model & Number of material constants & $R^{2}$ score \\
			\midrule
			Proposed \texttt{DSO} & 4 & $\SI{97.31578576913696}{\percent}$ \\
			Extended tube \citep{kaliske1999extended} & 4 & $\SI{96.56108483739724}{\percent}$ \\
			Non-hyperelastic Shariff \citep{Shariff2000} & 5 & $\SI{96.37595636432934}{\percent}$ \\
			Ogden model \citep{ogden1972large} & 6 & $\SI{95.58273523135516}{\percent}$\\
			\bottomrule
		\end{tabular}
\end{table}
The responses of these three models are illustrated in \autoref{fig:Treloar_Comparison}. Models with fewer number of parameters, such as the three-chain \citep{james1943theory}, Hart-Smith \citep{hart1966elasticity}, and eight-chain models \citep{Arruda1993} are unable to accurately predict the stress response over the entire strain range and are not used for comparisons with the proposed strain energy function. Note that according to \cite{Marckmann2006} for moderate strains of up to $200\SI{}{}$ - $250\SI{}{\percent}$, the two-parameter Mooney-Rivlin model demonstrates the greatest efficacy, exhibiting performance characteristics comparable to those of more complex models. For lower strains up to $150\SI{}{\percent}$, the neo-Hookean model is the preferred choice due to its physical basis, simplicity with a single parameter, and ability to predict material response in deformation modes. These conclusions are supported by the identified strain energy function  \eqref{equ:TreloarBestFitDSO}. The terms $\Psi_{1}$ and $\Psi_{2}$  serve as the fundamental building blocks for the neo-Hookean constitutive equation and the two-parameter Mooney-Rivlin model. As illustrated in \autoref{fig:TreloarContributionsZoom}, these contributions are most significant for the small and moderate strain ranges up to $250\SI{}{\percent}$. \\
In a comprehensive study \citep{ricker2023systematic} a range of hyperelastic models for nine distinct rubber compounds were compared and fitted to experimental results in addition to the classical Treloar data set. Specifically, the role and importance of the second principal invariant for rubber models were studied. Accordingly, UT is mainly influenced by $\IC$, while EBT and PS responses are equally affected by both $\IC$ and $\IIC$. Consequently, test data from a single experiment cannot adequately calibrate models dependent on both invariants.  Furthermore, $\IC$-based models results in an underestimation of the EBT response. By incorporating additionally $\IIC$, it is possible to achieve a balance in the stress response across diverse deformation modes and to offset potential limitations. These findings substantiate the observations made with the material model \eqref{equ:TreloarBestFitDSO} predicted by \texttt{DSO}. A significant benefit of the proposed methodology is that it does not necessitate an initial screening of any models, thereby preventing any potential bias in the model selection. Constitutive relations can be directly identified from the specified data and inputs.
\begin{table}[h]
	\centering
	\caption{Material parameters of the extended tube model \citep{kaliske1999extended} for the Treloar data set.}
	\label{tab:ExtendedTubeTreloarMaterialParam}
	\renewcommand{\arraystretch}{1.5}
	\begin{tabular}{cccc}
			\toprule
			$G_c$ & $G_e$ & $\beta$ & $\delta$ \\
			\midrule
			$\SI{0.19539293}{\mega \pascal}$  & $\SI{0.18874173}{\mega \pascal}$ & $\SI{0.33562464}{}$ & $\SI{0.09561381}{}$   \\
			\bottomrule
		\end{tabular}
\end{table}
\begin{table}[h]
	\centering
	\caption{Material parameters of the non-hyperelastic Shariff model \citep{Shariff2000} for the Treloar data set.}
	\label{tab:ShariffTreloarMaterialParam}
	\renewcommand{\arraystretch}{1.5}
	\begin{tabular}{cccccc}
			\toprule
			$E$ & $\alpha_{0}$ & $\alpha_{1}$ & $\alpha_{2}$ & $\alpha_{3}$   & $\alpha_{4}$ \\
			\midrule
			 $\SI{1.17089124e+00}{\mega \pascal}$  & $\SI{1.00e+00}{}$ & $\SI{8.64613504e-01}{}$ & $\SI{3.65921908e-02}{}$  & $\SI{8.35491892e-05}{}$ & $\SI{2.03900614e-02}{}$  \\
			\bottomrule
		\end{tabular}
\end{table}
\begin{table}[h]
	\centering
	\caption{Material parameters of the Ogden model \citep{ogden1972large} for the Treloar data set.}
	\label{tab:OgdenTreloarMaterialParam}
	\renewcommand{\arraystretch}{1.5}
	\begin{tabular}{cccc}
			\toprule
			& $i=1$ & $i=2$ & $i=3$  \\
			\midrule
			$\alpha_{i}$& $\SI{1.74016565e+00}{}$  & $\SI{7.27589533e+00}{}$ & $\SI{-1.82048513e+00}{}$\\
			$\mu_{i}$ & $\SI{4.13366411e-01}{\mega \pascal}$ & $\SI{1.21762075e-05}{\mega \pascal}$ & $\SI{-2.02988749e-02}{\mega \pascal}$ \\
			\bottomrule
		\end{tabular}
\end{table}
\clearpage
\subsection{Influence of Noise}
\begin{figure}[p]
	\centering
	\begin{subfigure}{0.5\textwidth}
			\includegraphics[width=1.2\textwidth]{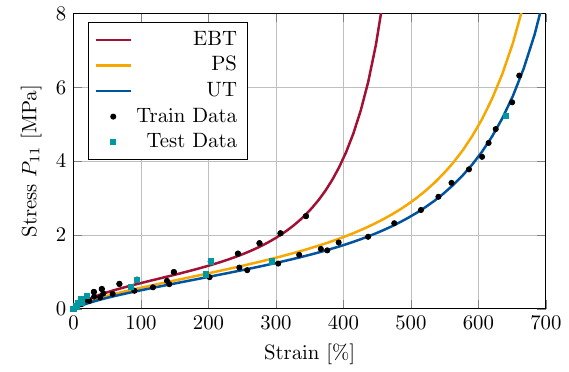}
			\caption{}
			\label{fig:TreloarFit_n1}
		\end{subfigure} \\
		\begin{subfigure}{0.5\textwidth}
			\includegraphics[width=1.2\textwidth]{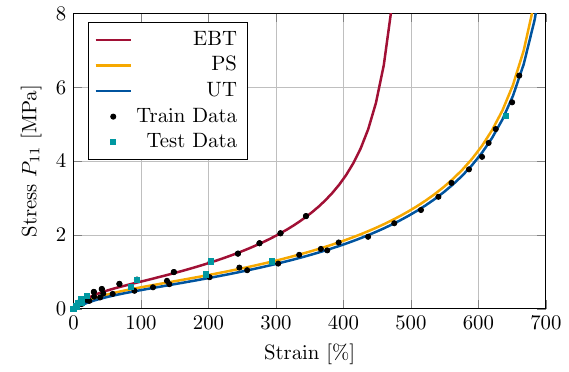}
			\caption{}
		 \label{fig:TreloarFit_n2}
		\end{subfigure}
	\caption[Stress-Strain Responses of Novel Models Under Different Noise Levels]{
			Stress-strain responses of strain energy functions \eqref{eq:Treloar_n1} and \eqref{eq:Treloar_n2} against the  experimental data by \cite{treloar1944stress} subject to two different levels of noise according to \eqref{eq:Noise}.}
	\label{fig:TreloarNoise}
\end{figure}

To study the quality of the data set provided and its impact on the performance of the obtained models we consider two distinct levels of noise 
\begin{align}
	n_{i}(\lambda) = a_{i} \frac{\lambda}{\lambda_{\text{max}}} \mathcal{N}(0, 1) \,  \label{eq:Noise}
\end{align}
imposed on the experimental data, where $\mathcal{N}(0, 1)$ represents the normal distribution, $a_{i}$ denotes the amplitude level, $\lambda$ and $\lambda_{\text{max}}$ are the current and the maximal stretch, respectively. Accordingly, the noise level is proportional to the amplitude of the current strain, and is normalized to the maximum stretch. \\ 
The aim of this study is to determine whether the derived strain energy function is a unique solution and to explore the robustness of the prediction in the presence of noise. 
We examined two distinct amplitude levels $a_1 = 0.025 \, \SI{}{\MPa}$ and $a_2 = \SI{0.05}{\MPa}$ and found the following strain energy functions, respectively
\begin{align}
	\Psi^{n_{1}}_{\text{T}} &= \frac{\ln{\IC} + \SI{28.42108203022402}{}}{- \SI{0.02770411058718299}{} \IC + \SI{1.1306629047538411}{} - \frac{1}{\IC}\SI{19.6374554685484}{}\exp\left(\SI{7.574960169390804}{} \exp\left(\SI{0.02064522578092095}{} \sqrt{\IC} - \frac{1}{\sqrt{\IIC}}\right) - \SI{16.150767419420287}{}\right)} \, , \label{eq:Treloar_n1} \\
	\Psi^{n_{2}}_{\text{T}} &= \SI{0.088520957704187183}{} \IC + \SI{0.10721797273102984 }{}\sqrt{\IIC} + \SI{0.10721797273102984 }{} \exp{\left(\SI{0.60023782905113407}{} \sqrt{\SI{0.74689298464165324 }{}\IIC - 1}\right)} + \exp\left(\SI{0.06724733103245549 }{}\exp\left[\exp\left(\SI{0.01729721736231109 }{}\IC\right)\right]\right) \, .
	\label{eq:Treloar_n2}
\end{align}
For illustrative purposes, the stress-strain response for both models is presented in \autoref{fig:TreloarNoise}. In particular, the strain energy function corresponding to the first noise level is observed to accurately describe both the PS and UT while the EBT response is slightly underestimated. Despite this discrepancy, the derived model captures the underlying data set with a high accuracy and a $R^{2}$ score of $\SI{98.19213091239288}{\percent}$. 
The second noise level formulation exhibits comparable patterns of behavior where the $R^{2}$ score is $\SI{97.14638855478787}{\percent}$. In this instance, the PS and UT responses demonstrate a high degree of similarity, while the EBT response is again underestimated. This may be attributed to the occurrence of data overlap within the strain range of $0\SI{}{\percent}$ to $200\SI{}{\percent}$, which is a direct result of the noise. Furthermore, this data overlap appears to result in a noticeably softer PS response in comparison to the first noise level for strains over $300\SI{}{\percent}$. \\
This investigation demonstrates the significance of accurate experimental data. Additionally, the results indicate multiplicity of solutions for the strain energy function capable to describe the underlying data set. While there is no guarantee of a function that can be decomposed additively, the derived models effectively and accurately fitted the provided stress-strain responses despite the sparsity of the available data.
Note also, that enforcing this additive decomposition in the strain energy is possible. However, it would significantly constrain the search space.
\subsection{Stretch-Based Model Identification}
\label{sec:RemarksStretchBasedTreloar}
\begin{figure}[p]
	\centering
	\includegraphics[width=0.6\textwidth]{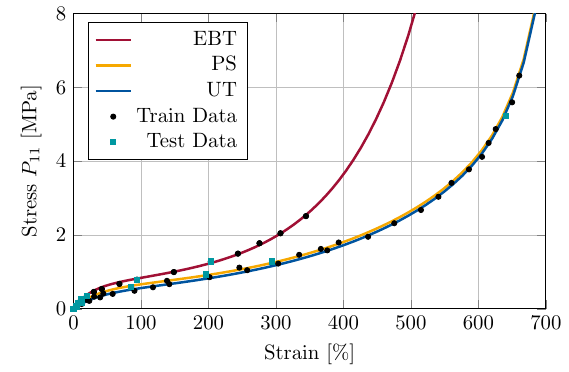}
	\caption[]{Stress-strain response of the stretch based model \eqref{eq:OgdenTreloar} against the Treloar data under UT, PS and EBT.}
	\label{fig:OgdenTreloar_1}
\end{figure}

Could an alternative stretch-based formulation provide a simpler expression with less material parameters or a superior fit to the underlying data set using \texttt{DSO}?  
To be able to find the Ogden model we extended the initial set of functions by the power one ("\textsf{pow}"). Once again, an $80\SI{}{\percent}$ to $20\SI{}{\percent}$  train-test split was employed. Within the Valanis-Landel concept \eqref{eq:ValanisLandelAssumption} the function $\omega$ was identified as follows:
\begin{align}
	&\omega(\lambda) = 1.44 \left(0.4\frac{2.96^{\lambda}}{\lambda} + \left(0.62 \sqrt{0.61 \lambda + \sqrt{\exp\left(\frac{2.0^{\lambda}}{\lambda}\right)}} - 1\right)^{0.65}\right)^{0.4} \, .\label{eq:OgdenTreloar} 
\end{align}
The stress-strain response resulting from this function is illustrated in \autoref{fig:OgdenTreloar_1} against the experimental data by Treloar. $R^{2}$ score of $\SI{98.0621}{\percent}$ indicates that the identified strain energy function accurately predicts the responses for all three loading cases and exhibits a superior performance compared to the Ogden model ($R^2 = \SI{95.58273523135516}{\percent}$).
This example demonstrates the effectiveness of a stretch-based approach. While accuracy improves, the notable increase in complexity of the model represents however a significant drawback. One potential avenue for enhancing the simplicity of this approach is the incorporation of possible priors into the \texttt{DSO} framework. Thus, further investigations and detailed analysis are necessary to fully determine the potential of this approach.
\clearpage
\subsection{Biaxial Loading of Vulcanized Rubber}
\label{sec:Kabwabata}
\begin{figure}[p]
	\centering
	\begin{subfigure}{0.5\textwidth}
			\includegraphics[width=1.2\textwidth]{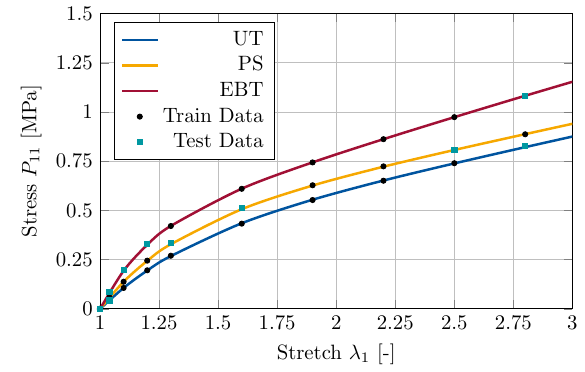}
			\caption{}
			\label{fig:Kabawata_P11_l1}
		\end{subfigure} \\
	\begin{subfigure}{0.5\textwidth}
		\includegraphics[width=1.2\textwidth]{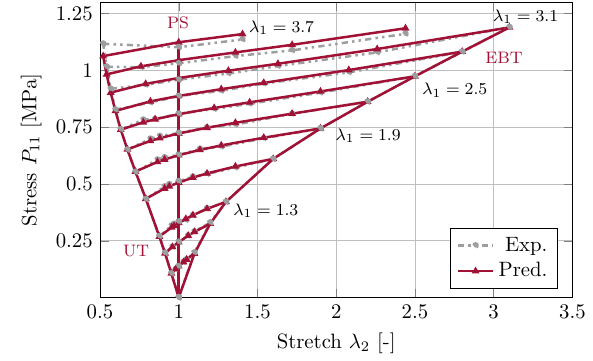}
		\caption{}
		\label{fig:Kabawata_P11_l2}
	\end{subfigure} \\
		\begin{subfigure}{0.5\textwidth}
			\includegraphics[width=1.2\textwidth]{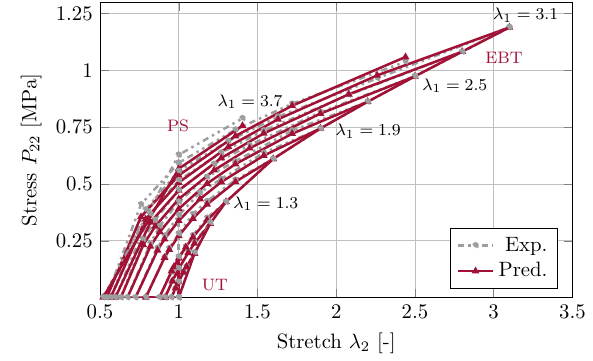} 
			\caption{}
		\label{fig:Kabawata_P22_l2}
		\end{subfigure} 
			\caption[]{
					Stress-stretch responses (\subref{fig:Kabawata_P11_l1}) $P_{11}(\lambda_{1})$ under UT,PS and EBT, (\subref{fig:Kabawata_P11_l2}) $P_{11}(\lambda_{2})$ and (\subref{fig:Kabawata_P22_l2}) $P_{22}(\lambda_{2})$ for different values of $\lambda_{11}$ plotted against the experimental data by \cite{kawabata1981experimental}. The responses result from the strain energy function \eqref{eq:Kawabata_1} identified on the basis of $P_{11}$ values from UT, PS and EBT.}
			\label{fig:Kabawata_1}
		\end{figure}

To evaluate the performance of the \texttt{DSO} package on the data set by \cite{kawabata1981experimental} from biaxial tension tests, three different training scenarios are explored. In the first scenario the UT, PS, and EBT responses are used to identify a strain energy function. The remaining data are used to evaluate the performance of the model, specifically its ability to predict the normal 1st Piola-Kirchhoff stresses in the loading directions 1 and 2 ($P_{11}$  and $P_{22}$). In the second scenario, all data sets for the biaxial tension tests with various stretch relations are used to to improve the performance of the model. The model is discovered only on the basis of $P_{11}$ stress values. In the final scenario, the entire data set is used to evaluate the ability of the model to accurately predict all data points including the $P_{11}$ and $P_{22}$ stresses. A test train split of $70\SI{}{\percent}$ to $30\SI{}{\percent}$ is applied for all three cases. Here, the strain energy function was determined as a function of the invariants $\IC$ and $\IIC$. \\
For the first example, the following strain energy is identified
\begin{align}
	\Psi_{\text{K},1} =  &\SI{0.43417905491527578}{} \sqrt{\Big(\SI{0.38280727876875269}{} \Big[\IC \Big( \SI{0.30468766496136507}{} \IC - \SI{8.581704823563822e-4} \IIC \ln{\IIC}- \SI{0.30468766496136507}{} \ln{\IC} + \SI{4.7327460737281793}{} \Big) +  \IIC\Big] + 1\Big)} \, . \label{eq:Kawabata_1}
\end{align}
It demonstrates a high degree of agreement with the experimental data for UT, PS and EBT with $R^{2}$ score of $\SI{99.66993806403691}{\percent}$ as depicted in \autoref{fig:Kabawata_1}\subref{fig:Kabawata_P11_l1}.
		Nevertheless, it is important to study how accurately the strain energy will predict responses in other scenarios. 
		
\begin{figure}[h!]
			\centering
				\begin{subfigure}{0.5\textwidth}
					\includegraphics[width=1.2\textwidth]{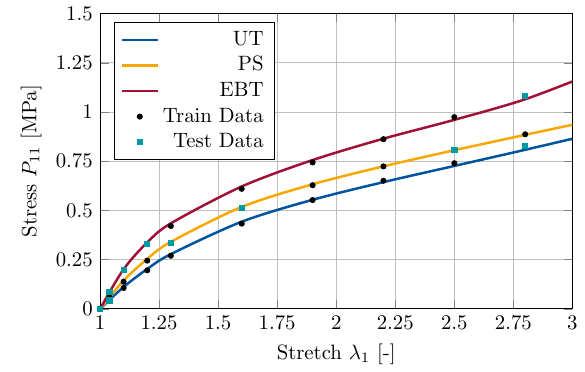}
					\caption{}
					\label{fig:Kabawata_3_P11_l1}
				\end{subfigure} \\
			\begin{subfigure}{0.5\textwidth}
					\includegraphics[width=1.2\textwidth]{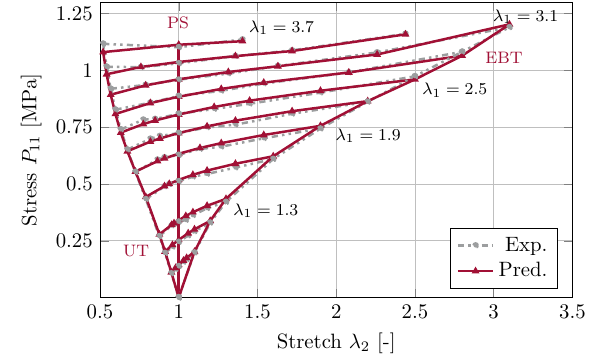}
					\caption{}
					\label{fig:Kabawata_3_P11_l2}
				\end{subfigure} \\
			\begin{subfigure}{0.5\textwidth}
					\includegraphics[width=1.2\textwidth]{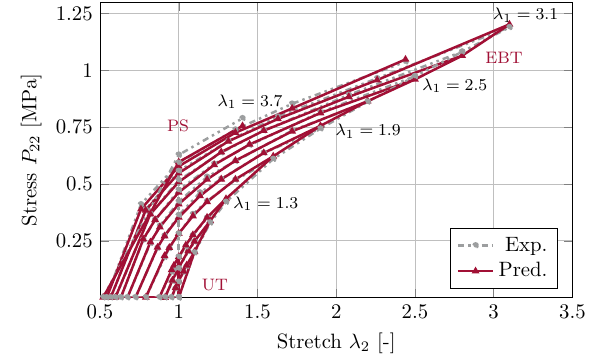} 
					\caption{}
					\label{fig:Kabawata_3_P22_l2}
				\end{subfigure} 
			\caption{Stress-stretch responses (\subref{fig:Kabawata_3_P11_l1}) $P_{11}(\lambda_{1})$ under UT,PS and EBT, (\subref{fig:Kabawata_3_P11_l2}) $P_{11}(\lambda_{2})$ and (\subref{fig:Kabawata_3_P22_l2}) $P_{22}(\lambda_{2})$ for different values of $\lambda_{11}$ plotted against the experimental data by \cite{kawabata1981experimental}. The responses result from the strain energy function \eqref{eq:Kawabata_3} identified on the basis of $P_{11}$ values from UT, PS and EBT.}
			\label{fig:Kabawata_3}
		\end{figure}

		In \autoref{fig:Kabawata_1}\subref{fig:Kabawata_P11_l2} and \autoref{fig:Kabawata_1}\subref{fig:Kabawata_P22_l2} the responses $P_{11} (\lambda_{1})$ and $P_{22} (\lambda_{2})$, respectively, are plotted for various values of $\lambda_1$.
		\autoref{fig:Kabawata_1}\subref{fig:Kabawata_P11_l2} demonstrates that these three loading cases are sufficient for characterizing the material behavior across various stretch combinations. However, $P_{22}$ response which was not seen by \texttt{DSO} only qualitatively describes the experimental data. Thus, more information is need for more accurate stress predictions in the biaxiaxial tension. \\
		For the second training case the following strain energy is identified
		\begin{align}
			\Psi_{\text{K},2} =& \SI{0.16702143336426614 }{}\IC - \SI{0.027744317551027826}{} +\SI{0.0958024930831093}{}  \Bigg(\SI{3.6946795686172456}{} - \SI{2.211859814364515}{} \sqrt{\SI{0.085072165716625369}{} \IC - \SI{0.2044018411354622}{} \sqrt{\IIC} + 1}\Bigg) \notag\\ &\cdot \ln{\left( \SI{4.7443898424407 }{} \IC + \IIC + \ln{\left(\sqrt{\IIC} \right)} + \frac{\SI{5.2132171735409315}{}}{\IC} \right)} \, . \label{eq:Kawabata_2}
		\end{align}
		It achieved a $R^{2}$ score of $\SI{99.4}{\percent}$.
		\begin{figure}[p]
			\centering	
			\begin{subfigure}{0.5\textwidth}
					\includegraphics[width=1.2\textwidth]{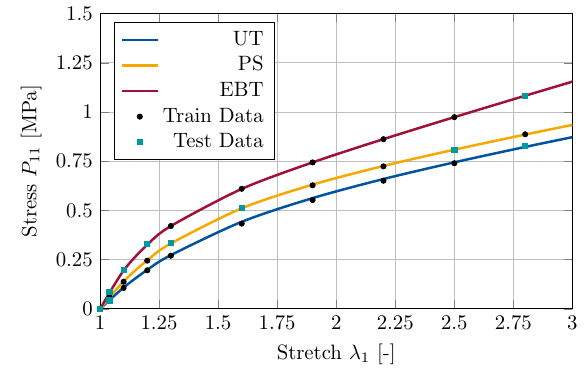}
					\caption{}
					\label{fig:Kabawata_2_P11_l1}
				\end{subfigure} \\
			\begin{subfigure}{0.5\textwidth}
					\includegraphics[width=1.2\textwidth]{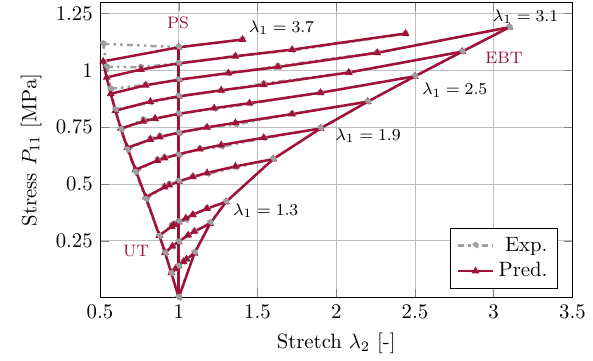}
					\caption{}
					\label{fig:Kabawata_2_P11_l2}
				\end{subfigure} \\
			\begin{subfigure}{0.5\textwidth}
					\includegraphics[width=1.2\textwidth]{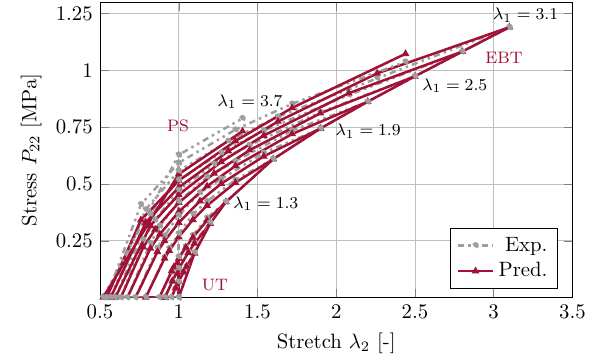} 
					\caption{}
					\label{fig:Kabawata_2_P22_l2}
				\end{subfigure} 
			\caption{Stress-stretch responses (\subref{fig:Kabawata_2_P11_l1}) $P_{11}(\lambda_{1})$ under UT,PS and EBT, (\subref{fig:Kabawata_2_P11_l2}) $P_{11}(\lambda_{2})$ and (\subref{fig:Kabawata_2_P22_l2}) $P_{22}(\lambda_{2})$ for different values of $\lambda_{11}$ plotted against the experimental data by \cite{kawabata1981experimental}. The responses result from the strain energy function \eqref{eq:Kawabata_2} identified on the basis of $P_{11}$ values from UT, PS and EBT.}
			\label{fig:Kabawata_2}
		\end{figure}
		In this case, all data from the biaxial loading were utilized. However, only the $P_{11}$ response was fitted, while the $P_{22}$ response was excluded from the fitting process. The corresponding stress-strain responses are illustrated in \autoref{fig:Kabawata_2} against the experimental data by \cite{kawabata1981experimental}. It is seen that the predictions based on \eqref{eq:Kawabata_2} are very accurate for UT, PS and BT, effectively capturing all observed trends. However, the predictive accuracy of the model for $P_{22}$ is less precise in comparison to $P_{11}$. In particular, $P_{22}$ is considerably underestimated for $\lambda_1=3.7$.  This indicates that both stresses are indispensable for the fitting. \\
		In the final case, both $P_{11}$ and $P_{22}$ responses were used for the fitting. The strain energy was determined with
		\begin{align}
			\Psi_{\text{K},3} =&  \SI{0.12223415594624147 }{}\IC + \SI{0.12223415594624147 }{}\sqrt{\IIC} - \SI{0.025169754009035435}{} \ln{\IIC}  - \SI{0.025169754009035435}{} \ln{\left(\ln{\left(\IIC \right)}^{4} \right)} - \SI{0.12995614747379221}{} \notag\\ & - \frac{\SI{0.025169754009035435}{} \left(- \SI{3.4926060582710092}{} \IC - \SI{5.8774593591164961}{} \IIC\right) \ln{\left(\IIC \right)}}{\IIC} \, . \label{eq:Kawabata_3}
		\end{align}
		The resulting responses are visualized in \autoref{fig:Kabawata_3}. In this instance, the generated predictions are observed to be of an extremely high degree of accuracy with a $R^{2}$ score of $\SI{98.4}{\percent}$. Nevertheless, due to the larger number of data points used compared to the second case, the accuracy is slightly lower.
		The fitting of both stress responses $P_{11}$ and $P_{22}$ has resulted in a notable enhancement in the precision of the calculated $P_{22}$ response. \\
		This analysis indicates that there is no single, universally applicable strain energy function, rather, multiple potential functions can adequately fit the same data set. The identified strain energy functions illustrate that both $\IC$ and $\IIC$, are indispensable as arguments of the strain energy function. In order to obtain a precise model capable of predicting a variety of loading scenarios, both the $P_{11}$ and $P_{22}$ responses are indispensable for the fitting process. While the experimental data obtained from UT, PS, and EBT experiments can yield satisfactory fits, it is crucial to acknowledge that these data may not provide highly accurate predictions for specific loading cases, particularly with regard to the $P_{22}$ response. The $P_{11}$ stress is sufficiently robust for the identification of a strain energy function which effectively captures the observed trends in the $P_{22}$ response.
		\clearpage
		\section{Conclusion}
		\label{sec:Conclusion}
		In this study, we have proposed a novel methodology that employs deep symbolic regression to derive interpretable hyperelastic material models for rubber-like materials under multi-axial loading conditions. By directly utilizing experimental data from the classical Treloar and Kawabata data sets, our methodology effectively identifies strain energy functions that not only accurately fit the data but also require few material parameters. This approach has been subjected to rigorous testing for both invariant and stretch-based formulations using UT, PS and EBT data. In conclusion, the present study demonstrates the considerable potential of integrating deep symbolic regression with a continuum mechanical framework, even when data is limited. The important advantage of the proposed procedure is that it bypasses human bias in the model selection, thus more effectively capturing the complex behavior of rubber-like materials. Furthermore, the proposed methodology demonstrates robustness against data noise and versatility in predicting responses across various loading scenarios. The symbolic character of the identified model allows to analyze and interpret the contribution of each term under different deformation modes. Domain-specific priors incorporated into the DSO framework will further be able to restrict the search space and simplify the resulting models.

\bibliographystyle{cas-model2-names}

\bibliography{cas-refs}

\end{document}